\documentclass[conference]{IEEEtran}
\IEEEoverridecommandlockouts

\usepackage{flushend}

\usepackage[utf8]{inputenc}
\usepackage[english]{babel}
\usepackage{microtype}
\usepackage{csquotes}
\renewcommand{\enquote}[1]{\textit{#1}}

\usepackage[
  backend=biber,
  citestyle=ieee,
  bibstyle=ieee,
  url=false,
  isbn=false,
  eprint=false,
]{biblatex}
\usepackage[nohyperlinks, printonlyused, nolist]{acronym}
\begin{acronym}
  \acro{SMC}{semi-Markov chain}
  \acro{WCET}{worst-case execution time}
  \acro{GEV}{generalised extreme value}
  \acro{GMM}{Gaussian mixture model}
  \acro{CPU}{central processing unit}
  \acro{EM}{expectation-maximisation}
  \acro{AIC}{Akaike information criterion}
  \acro{BIC}{Bayesian information criterion}
  \acro{HMM}{hidden Markov models}
  \acro{OS}{operating system}
  \acroplural{OS}[OSes]{operating systems}
  \acro{RTOS}{real-time operating system}
  \acro{YAML}{YAML Ain't Markup Language}
  \acro{CSV}{comma-separated values}
  \acro{MBPTA}{measurement-based probabilistic timing analyse}
  \acro{SPTA}{static probabilistic timing analysis}
  \acro{STA}{static timing analysis}
  \acro{BBM}{basic block measurement}
  \acro{MLE}{maximum likelihood estimation}
  \acro{BN}{Bayesian network}
  \acro{DAG}{directed acyclic graph}
  \acro{SMM}{semi-Markov model}
  \acro{HSMM}{hidden semi-Markov model}
  \acro{CDF}{cumulative distribution function}
  \acro{BMM}{block maximum method}
  \acro{PoT}{peak over threshold}
  \acro{SBC}{single-board computer}
  \acro{TTP}{timed tracepoint}
  \acro{EVT}{extreme value theory}
  \acro{ID}{identifier}
  \acro{IID}{independent and identically distributed}
  \acro{ET}{execution time}
  \acro{pET}{probabilistic execution time}
  \acro{pWCET}{probabilistic worst-case execution time}
  \acro{FIFO}{first in first out}
  \acro{EVA}{extreme value analysis}
  \acro{AST}{abstract syntax tree}
  \acro{DFA}{deterministic finite automaton}
  \acroplural{DFA}[DFAs]{deterministic finite automata}
\end{acronym}

\acused{CPU}
\acused{ID}
\acused{OS}
\acused{YAML}
\acused{CSV}
\acused{FIFO}

\usepackage[intlimits]{mathtools}
\usepackage{amsthm, amssymb, amsfonts, amsbsy}
\usepackage{nicefrac}
\usepackage[
  colorlinks=true,urlcolor=teal,linkcolor=teal,citecolor=teal,
  final=true,
]{hyperref}
\hypersetup{unicode=true}
\usepackage{tabularray}
\UseTblrLibrary{booktabs}
\usepackage[dvipsnames]{xcolor}
\usepackage[
  group-digits = integer,
  group-minimum-digits={3},
  reset-text-family = false,
  reset-text-series = false,
  reset-text-shape = false,
  propagate-math-font = true,
  text-series-to-math = true,
]{siunitx}

\definecolor{lfd2}{HTML}{E69F00}
\definecolor{lfd3}{HTML}{999999}
\definecolor{lfd4}{HTML}{009371}
\definecolor{lfd5}{HTML}{BEAED4}
\definecolor{lfd6}{HTML}{ED665A}
\definecolor{lfd7}{HTML}{1F78B4}

\newcommand*{\ie}{\textit{i.e.}}
\newcommand*{\eg}{\textit{e.g.}}
\newcommand*{\etc}{etc\@ifnextchar.{}{.\@}}
\newcommand{\circledinline}[1]{\raisebox{.9pt}{\textcircled{\raisebox{-.9pt}{#1}}}}
\addto\extrasenglish{%
}

\begin{document}

\title{
  From Tracepoints to Timeliness: A Semi-Markov Framework for Predictive Runtime Analysis
  \thanks{%
    This work was supported by the European Union (Project Reference 101083427) and the European Funds for Regional Development~(EFRE) (Project Reference 20-3092.10-THD-105).
  }
}

\author{
  \IEEEauthorblockN{Benno Bielmeier}
  \IEEEauthorblockA{\textit{Technical University of} \\
    \textit{Applied Sciences Regensburg} \\
    Regensburg, Germany \\
  \href{mailto:benno.bielmeier@othr.de}{benno.bielmeier@othr.de}}
  \and
  \IEEEauthorblockN{Ralf Ramsauer}
  \IEEEauthorblockA{\textit{Technical University of} \\
    \textit{Applied Sciences Regensburg} \\
    Regensburg, Germany \\
  \href{mailto:ralf.ramsauer@othr.de}{ralf.ramsauer@othr.de}}
  \and
  \IEEEauthorblockN{Takahiro Yoshida}
  \IEEEauthorblockA{
    \textit{Tokyo University of Science} \\
    Tokyo, Japan\\
  \href{mailto:yoshida@ee.kagu.tus.ac.jp}{yoshida@ee.kagu.tus.ac.jp}}
  \and
  \IEEEauthorblockN{Wolfgang Mauerer}
  \IEEEauthorblockA{\textit{Technical University of}\\
    \textit{Applied Sciences Regensburg} \\
    \textit{Siemens AG, Technology} \\
    Regensburg/Munich, Germany \\
  \href{mailto:wolfgang.mauerer@othr.de}{wolfgang.mauerer@othr.de}}
}

\maketitle

\newcommand{\aca}[1]{\acl*{#1}}
\newcommand{\acap}[1]{\aclp*{#1}}

\begin{abstract}
  Detecting and resolving violations of temporal constraints in real-time systems is both, time-consuming and resource-intensive, particularly in complex software environments.
  Measurement-based approaches are widely used during development, but often are unable to deliver reliable predictions with limited data.

  This paper presents a hybrid method for \acl{WCET} estimation, combining lightweight runtime tracing with probabilistic modelling.
  Timestamped system events are used to construct a \acl{SMC}, where
  transitions represent empirically observed timing between events.
  Execution duration is interpreted as time-to-absorption in the \acl{SMC}, enabling \acl{WCET} estimation with fewer assumptions and reduced overhead.

  Empirical results from real-time Linux systems indicate that the method captures both regular and extreme timing behaviours accurately, even from short observation periods.
  The model supports holistic, low-intrusion analysis across system layers and remains interpretable and adaptable for practical use.
\end{abstract}


\section{Introduction}

Real-time systems with complex software stacks, comprising multiple interdependent layers and components, pose challenges for traditional \ac{WCET} analysis~\cite{davis_survey_2019}.
\Acl{STA} provides sound and conservative upper bounds with formal guarantees, making it indispensable for safety-critical systems.
However, it tends to yield overly pessimistic estimates and requires impractical effort, particularly when confronted with rare timing anomalies and context-dependent behaviours~\cite{becker_scalable_2019,meng_reducing_2017,graydon_realistic_2014}.

This challenge is further exacerbated in emerging high-performance and heterogeneous architectures, such as quantum-accelerated systems, where latency predictability and execution-time characterisation are critical but poorly understood at system level~\cite{Wintersperger2022}.
Similarly, modern hardware platforms employing mechanisms such as static hardware partitioning on RISC-V introduce architectural features that complicate timing analysis due to increased concurrency, resource contention, and architectural opacity~\cite{Ramsauer2022}.
The need for modular, transparent, and verifiable development processes, as exemplified in open source engineering contexts, underscores the importance of reproducibility and empirical validation in timing analysis~\cite{Mauerer2013}.

Probabilistic timing analysis techniques based on measurement data have emerged to address both issues by offering probabilistic estimates instead of deterministic timing guarantees~\cite{davis_survey_2019,jimenez_gil_open_2017, davis_survey_2019}.
Due to their inherent uncertainty, these methods are not meant as alternatives for deterministic methods in (safety) critical domains.
They rather provide valuable insights when firm guarantees are not (yet) required, for instance in soft real-time applications, or during development phases of a system.
They allow for rapid runtime validation, while avoiding extensive overheads known to be associated with more sophisticated exhaustive analyses.
However, existing \ac{MBPTA} approaches typically require large \ac{IID} sample sets and assume statistical independence, which can be difficult to ensure on complex hardware and leads to high measurement overheads.

\begin{figure*}[htb]
  \centering
  \includegraphics{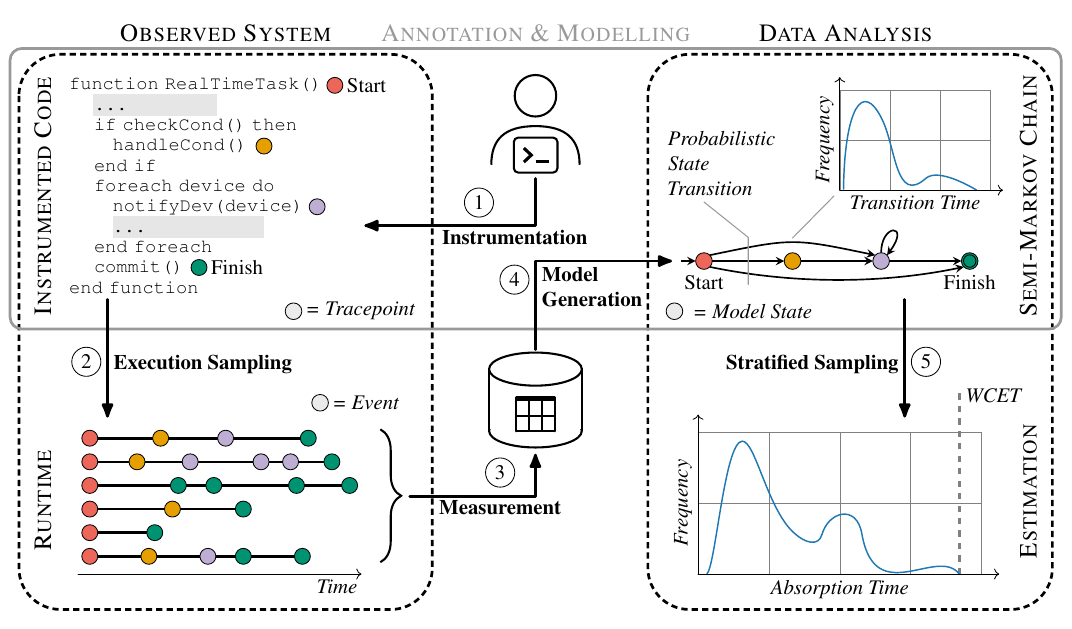}
  \caption{
    Conceptual overview of our approach.
    Based on expert knowledge about the architecture of a system under test, our  starts with system instrumentation, by enabling existing or inserted tracepoitns~\circledinline{1}.
    We execute the instrumented code repeatedly~\circledinline{2} and record the trace data, which we store as tabular data for offline analysis~\circledinline{3}.
    We utilise this data to generate a stochastic model in the form of an \acl{SMC}~\circledinline{4}.
    We analyse the model through sampling by conducting multiple simulation runs~\circledinline{5}.
    This analysis yields an estimated \acl{WCET}.
  }
  \label{fig:overview}
\end{figure*}

We propose the use of \acp{SMC} to stochastically model the probabilistic execution time of real-time tasks based on a limited number of runtime measurements.
This enables efficient timing analysis without requiring extensive system instrumentation, detailed manual modelling, or exhaustive trace evaluation.
The resulting model captures both the temporal characteristics---via stochastic transition delays---and the logical execution flow---through the structure of states and transitions.
This state-based representation facilitates not only intuitive visualisation and interpretation of distinct execution phases, but also supports flexible refinement guided by domain-specific knowledge of the target system.
By combining lightweight measurement with probabilistic modelling, the approach yields practical and scalable insights into timing behaviour, suitable even in early development stages or in settings with restricted observability.

Given a real-time task whose timing behaviour is to be analysed, our approach proceeds as outlined in \autoref{fig:overview}.
First, the target system is instrumented~\circledinline{1} to capture essential timing events using lightweight tracepoints;
these can appear, for instance, in task scheduling, during interrupt processing, or in function calls.
The placement of tracepoints is guided by system-specific considerations and expert judgement, depending on the concrete system, application, and analysed tasks, as detailed in \autoref{sec:method}.
Next, the system is executed under realistic conditions~\circledinline{2}, and trace events are recorded for subsequent analysis~\circledinline{3}.
This measurement data then serves as foundation to construct \acl{SMC} models~\circledinline{4}.
The probability distribution of \ac{pWCET} is derived by simulating numerous state transition sequences~\circledinline{5}, where each sequence represents the execution of a single task run.
The resulting distribution of  \aclp{ET} is then analysed to extract probabilistic temporal properties, such as quantile bounds and estimates of the \ac{WCET}.
These values characterise the task's temporal behaviour under realistic conditions and can inform design decisions, safety arguments, and schedulability assessments in real-time systems.

By constructing \ac{SMC} models exclusively from trace points in all software layers of the system (\ie, kernel space, user space, and possibly firmware), our framework is holistically applicable to the whole software stack.
It can seamlessly integrate data from established tracing mechanisms or custom implementations, underscoring its practical flexibility.
Rather than isolated or individual reasoning on single
observations, our method probabilistically combines effects of scheduling decisions, hardware influence, and other systemic effects, and thereby simplifies abstract analysis of complex real-world systems.

Moreover, this hybrid approach---positioned between rigorous formal verification and purely empirical analysis---not only mitigates the inherent pessimism of traditional deterministic \ac{WCET} methods, but also reduces the complexity and rigidity typically associated with such methods~\cite{jimenez_gil_open_2017,santinelli_revising_2017}.
An important practical benefit is to assist developers, particularly as the framework readily provides rapid feedback on the effect of code changes to real-time behaviour, similar to the goal of DevOps efforts.
This allows for early identification of performance regressions or improvements, and thereby accelerated development cycles and facilitates targeted optimisations.
The low overhead of our measurement technique further supports integration into continuous integration pipelines to provide early feedback to developers.

Our main contributions are
\begin{itemize}
  \item
    a hybrid analysis method that integrates measurement-based observations with probabilistic \acl*{SMC} modelling for real-time system analysis;

  \item
    a demonstration of real-world applicability and the effectiveness of runtime modelling with an evaluation of worst-case latency prediction, measured by cyclictest, a standard Linux RT test suite, on a RISC\nobreakdashes-V target;

  \item
    an assessment of the model's quality with respect to the amount of measurement data to build expressive models.

\end{itemize}

This paper is structured as follows.
\autoref{sec:backgrounds} provides background information on timing analysis of real-time system and \acl{SMC}.
After the presentation of related work in \autoref{sec:related-work}, we detail our method in \autoref{sec:method}.
In \autoref{sec:implementation} we present the implementation components of the approach.
\autoref{sec:evaluation} evaluates the approach for worst-case latency estimation of cyclictest on a real-time Linux system.
We discuss our results in \autoref{sec:discussion} and conclude, together with suggestions for future research,
in \autoref{sec:conclusion}.

\section{Background}%
\label{sec:backgrounds}

\subsection{Timing Analysis in Real-Time Systems}

In real-time computing systems, correctness depends on both, correct functional, and precise temporal behaviour.
Payloads (\eg, \textit{real-time} tasks) must meet strict timing constraints.
Missed deadlines may lead to severe consequences.
Such tasks often exhibit event-driven behaviour, that is, they are initiated or triggered by specific event occurrence or changes in system state.
Determining a task's \ac{WCET}---the maximum execution time a task can exhibit under any feasible scenario---is central to ensure these constraints~\cite{abella_wcet_2015,bernat_wcet_2002}.

Traditionally, \ac{WCET} analysis relies on either \textit{static} or \textit{measurement-based} approaches~\cite{abella_wcet_2015}.
Static approaches that employ modelling and formal verification provide safe bounds, but are often overly conservative owing to the complexity of accurately modelling advanced hardware features such as caches, pipelines, and multicore interaction bounds~\cite{abella_wcet_2015}.
They also do not apply well to complex real-world systems like Linux, for reasons ranging from state explosion to
practically unmanageable computational  requirements~\cite{goos_model_1998,jeon_quantum_2024}.

Measurement-based approaches estimate \ac{WCET} from empirical data obtained through execution profiling.
They offer more realistic but non-exhaustive estimates that risk underestimating worst-case conditions.
\textit{Hybrid} methods attempt to balance these limitations by combining elements of static and empirical analyses~\cite{betts_hybrid_2010,edgar_statistical_2001}.

The mentioned challenges with excessive pessimism or insufficient coverage motivated the transition towards probabilistic timing analysis~\cite{davis_survey_2019,maiza_survey_2019}.
Probabilistic methods, notably \ac{SPTA} and \ac{MBPTA}, characterise execution times as probability distributions, and thus explicitly model uncertainty~\cite{bernat_wcet_2002}.
While \ac{SPTA} analytically computes execution-time probabilities from program and hardware models, \ac{MBPTA} statistically derives execution-time distributions from measurements, and extrapolates probabilistic \ac{WCET}.
Although probabilistic techniques significantly reduce pessimism, ensuring validity requires careful assumptions about hardware and execution conditions.

\subsection{Observation of Real-Time Systems}

Reliable observation of real-time system behaviour requires accurate and minimally intrusive event capture.
While hardware-assisted tracing offers precise timestamps with negligible perturbation, it depends on specialised equipment, increases cost, and often lacks visibility into higher software layers.
Software-based tracing has become widespread due to its flexibility across system layers, lack of hardware requirements, and support for fine-grained, customisable instrumentation.
However, it can introduce runtime overhead and perturb timing behaviour, potentially distorting observed phenomena.
Tracing methods thus span software, hardware-assisted, and hybrid approaches~\cite{brandenburg_comparison_2013,sharma_hardwareassisted_2016,bao_hmtt_2011,de_oliveira_timing_2016}, each balancing accuracy, intrusiveness, and implementation effort~\cite{hofmann_distributed_1994,verge_hardwareassisted_2017}.

An essential requirement for tracing real-time systems is to maintain minimal overhead, as any perturbation degrades analysis quality.
Lightweight instrumentation is therefore critical to preserve runtime characteristics.
Efficient data collection methods, such as memory-resident ring buffers or selective logging, help ensure low overhead and practical applicability in real-time contexts.

Trace data may be processed offline or online.
Offline analysis collects events for later evaluation, offering flexibility and low hardware demands, albeit with delayed insights.
Online analysis, in contrast, processes events during execution for immediate feedback and low memory usage, but is more complex and may require significant computational resources~\cite{cheng_real-time_2002}.

Linux-based systems benefit from a mature ecosystem of tracing tools.
Prominent frameworks such as \texttt{ftrace}, \texttt{eBPF}, and \texttt{LTTng} support efficient in-memory recording, and both offline and online analysis modes.
However, their general-purpose design and associated instrumentation overhead can interfere with the timing behaviour of real-time tasks, particularly when analysing latencies in the sub-microsecond range.
This limits their applicability in strict real-time scenarios and often necessitates the development of specialised, low-intrusion tracing solutions tailored to the specific characteristics and constraints of the target system.

\subsection{Semi-Markov Chain}

\begin{figure}
  \centering
  \includegraphics{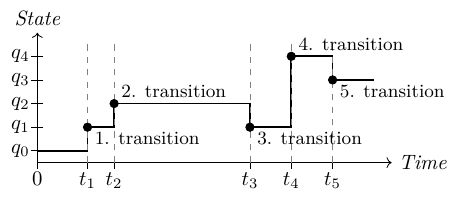}
  \caption[Visualisation of a semi-Markov Chain Trajectory]{%
    Trajectory of a \acl{SMC}, visualising one realisation of its state evolution over time.
    The horizontal axis represents continuous time~$\mathcal{T}$, while the vertical axis corresponds to the discrete set of states~$\mathcal{Q}$.
    The trajectory starts with initial state~$J_0 = q_0$ and transitions at time~$t_1$ to state~$q_1$.
    Assuming no state change happens after $t_5$, $q_3$ is considered an absorbing state.
  }
  \label{fig:smc}
\end{figure}

A \acfi{SMC} is a stochastic process $Z(t)_{t \in \mathcal{T}}$ defined over a finite state space~$\mathcal{Q}$ and an index set~$\mathcal{T}$, representing time (discrete or continuous).
It models the system's temporal evolution as a trajectory of state transitions, as visualised in \autoref{fig:smc}, which shows the system's state at each time.
In contrast to standard Markov chains, \acp{SMC} permit arbitrary sojourn time distributions between transitions, thus enabling the modelling of non-Markovian, memory-dependent behaviour~\cite{vassiliou_markov_2021,barbu_semi-markov_2008}.
In particular, the time spent in a state need not follow the exponential distribution, but may also depend on the elapsed time since the last transition.

Formally, an \ac{SMC} is defined as a sequence of state-time pairs $(J_n, T_n)_{n \in \mathbb{N}}$, where $J_n \in \mathcal{Q}$ denotes the state entered at time $T_n \in \mathcal{T}$, and the sequence $(T_n)$ is strictly increasing.
The time spent in state $J_n$ before transitioning is the sojourn time $S_n = \nolinebreak T_{n+1} - T_n$, characterised by the cumulative distribution function
$F_i(d) = \Pr(S_n \leq d \mid J_n = q_i)$
where $q_i \in \mathcal{Q}$ and $d \in \nolinebreak \mathcal{T}$.
State transitions are governed by the probability matrix $P$ with entries
$P_{ij} = \Pr(J_{n+1} = q_j \mid J_n = q_i)$
where $q_i, q_j \in\nolinebreak\mathcal{Q}$
subject to $\sum_j P_{ij} = 1$ for all $q_i$.
We consider a transition $q_i \to q_j$ absent if $P_{ij} = 0$.

The joint distribution of transitioning from $q_i$ to $q_j$ within a duration $d \in \mathcal{T}$ is given by the conditional sojourn time distribution
$F_{ij}(d) = \Pr(S_n \leq d, J_{n+1} = q_j \mid J_n = q_i)$
which forms the hold time kernel $Q$.

The initial distribution $\pi$ assigns probabilities to initial states in which the \ac{SMC} starts at $t=0$, satisfying $\pi_i = \Pr(J_0 = q_i)$ with $\sum_{i=1} \pi_i = 1$.
We denote the set of states with non-zero initial probability as $\mathcal{S} = \nolinebreak \{ q_i \in \mathcal{Q} \mid \pi_i > 0 \}$.

We assume time-homogeneity, meaning that the transition probabilities $P_{ij}$ and sojourn time distributions $F_{ij}$ are independent of both absolute time and the transition index $n$.

A state $q_i$ is considered \emph{absorbing} if $P_{ij} = 0$ for all $j \neq i$, equivalently $P_{ii} = 1$.
The set of absorbing states is defined as
$\mathcal{A} = \{ q_i \in \mathcal{Q} \mid P_{ii} = 1 \}$.
The process is \emph{absorbed} once it enters any $q_a \in \mathcal{A}$.
Let
$N = \min\{ n \ge 0 \mid J_n \in \mathcal{A} \}$
be the index of the absorbing transition, which is also the length of the process' state sequence.
The corresponding \emph{time to absorption} is then
$T_{\text{abs}} = T_N = \sum_{n=0}^{N-1} S_n$.

\section{Related Work}%
\label{sec:related-work}

\Acp{SMM} are widely used to capture systems with memory-dependent temporal behaviour.
Applications span reliability analysis~\cite{damico_reliability_2013}, predictive maintenance~\cite{cartella_hidden_2015}, biological sequence classification~\cite{ruiz-suarez_hidden_2022,barbu_semi-markov_2008}, and pattern recognition such as speech and handwriting~\cite{yu_hidden_2010}.
These models allow arbitrary sojourn time distributions and are often extended with hidden states to represent latent dynamics.

In real-time systems, Bozhko et al.~\cite{bozhko_what_2023} introduce the first axiomatic characterisation of \acf{pWCET}.
Their Coq-verified framework establishes a mathematically sound upper bound on the execution time distribution under \ac{IID} assumptions.
However, the model is abstract, defers empirical validation, and presumes statistical independence once \acp{pWCET} are derived.
Our approach contrasts with this by constructing empirical semi-Markov models from lightweight tracepoints, capturing system-wide timing behaviour without relying on the \ac{IID} assumption.
As detailed on in \autoref{sec:method}, execution time is modelled as time-to-absorption, using \acp{GMM} and stratified sampling to estimate \ac{pWCET}.
Thus, while Bozhko et al. provide formal soundness, our method offers a scalable, tool-supported alternative suited to data-constrained scenarios.

Several state-based \ac{MBPTA} approaches share conceptual ground with our method.
In particular, \emph{Basic Block Measurements}, \emph{Bayesian Networks}, and \emph{Timed Automata} have been used as probabilistic and state-based methods for runtime analysis.

Basic block measurement is a hybrid approach used in program profiling and performance analysis~\cite{chennupati_machine_2021,graydon_realistic_2014}.
It identifies execution hotspots, optimisation opportunities, or security vulnerabilities at the \ac{CPU} level~\cite{thoma_basicblocker_2021,ragel_secured_2015}.
A program is decomposed into \emph{basic blocks}, i.e., straight-line code sequences with a single entry and exit point, for which metrics such as execution count, time, or resource usage are determined~\cite{chennupati_machine_2021,lin_precise_2021}.
However, basic block-based techniques generally scale poorly to large code bases comprising millions of blocks, and they are ill-suited for analysing parallel, multithreaded, or highly dynamic programs.
They rely on a fixed control-flow structure, rendering them vulnerable in the context of dynamic code generation, self-modifying behaviour, or even common compiler optimisations such as inlining or loop unrolling.
To address some of these limitations, the WE-HML approach~\cite{amalou_we-hml_2021} combines static control-flow analysis with machine learning-based \ac{WCET} estimation at the binary level.
It mitigates the impact of compiler rearrangements by working on compiled code and models non-deterministic timing effects caused by caches via learned pollution-sensitive timing models.
By training on a large set of auto-generated code fragments, WE-HML improves coverage and generalisation while avoiding per-application measurements.

\Aclp{BN} are state-based probabilistic models in which every node represents a random variable and the edges denote conditional dependencies between them.
Their strength lies in modelling uncertainty, interference capability, and probabilistic reasoning, enabling the estimation of confidence intervals for \ac{WCET}~\cite{cai_application_2019}.
Applied in the field of runtime analysis, they focus on modelling complex probabilistic dependencies among a set of variables using \aclp{DAG}~\cite{zhang_bayesian_2023,marcot_advances_2019}.
While \acp{SMC} model temporal behaviour explicitly through sojourn time distributions, Bayesian Networks model the transition times implicit through dependencies among variables~\cite{puga_bayesian_2015}.
Bayesian Networks are well suited when dealing with complex (runtime) dependencies and requiring probabilistic inference to handle uncertainties in system behaviour.
However, their reliance on static conditional probabilities may limit the accurate capture of dynamic temporal behaviours, particularly in systems where timing plays a critical role.
Moreover, the computational complexity of probabilistic inference in large-scale networks remains a challenge.

State-based models are foundational not only for timing analysis but also for formal verification of functional and temporal properties.
Finite-state machines, such as \acp{DFA}, specify valid event sequences, while timed automata extend them with real-valued clocks to capture real-time constraints~\cite{halbwachs_timed_1999}.
These models support runtime verification via model checking and online monitoring.
Oliveira et al.~\cite{de_oliveira_thread_2020,de_oliveira_efficient_2019,de_oliveira_automata-based_2020}, for example, employ automata-based monitors to verify that events occur only in valid temporal and causal contexts.
Though these approaches focus on correctness rather than timing estimation, they demonstrate the expressiveness of state-based models.
Our semi-Markov framework complements them by adding stochastic timing semantics, enabling probabilistic analysis within a unified, state-based abstraction.

Lesange et al.~\cite{lesage_framework_2015} proposed a framework for measurement-based timing analyses, using an abstract model of synthetic tasks, derived via dynamic instrumentation using Valgrind on FFmpeg.
Execution times are aggregated via discrete convolution over syntax-tree-like structures with typed nodes (\eg, loop, conditional), resembling basic block methods.
Although this fine-grained abstraction captures specific execution characteristics, it does not constitute dynamic behaviour, flexibility, and the option to extrapolate unobserved behaviours (on possible unseen paths) like our method allows utilizing a semi-Markov framework with a stochastic process.

Friebe et al.~\cite{friebe_continuous-emission_2023} model execution times using Markov chains with Gaussian emissions to bound deadline-miss probabilities in reservation servers.
To avoid exponential sojourn times, they introduce hidden states.
However, their method assumes regularity and is less effective in capturing non-deterministic phenomena such as interrupts.
Our semi-Markov approach directly models such variability with mathematically grounded support for complex timing behaviour and minimal measurement intrusion.

\section{Method}%
\label{sec:method}

Our approach employs \acp{SMC} to capture the probabilistic runtime behaviour of tasks, explicitly accounting for both intrinsic task execution and non-deterministic interference from concurrent system activities.
We construct an abstract probabilistic model based on runtime measurements.
It aims to characterise the distribution of latency values, with particular emphasis on sparsely populated regions where extrapolation is challenging, such as the tail of the latency distribution.

This framework reflects the idea that the execution of a real-time task can be conceptualised as a finite sequence of distinct segments separated by certain events or conditions, for example:
task initiation, execution, preemption, resumption, and termination.
The duration of each segment is influenced by various factors, such as input data characteristics, other system activities, and microarchitectural state of the hardware.
An \ac{SMC} naturally accommodates this scenario and represents each segment as a state, with the hold-time distribution estimated from the timestamps of observed events.

Unlike methods that assume \acf{IID} execution times, our \ac{SMC} approach retains context by conditioning the timing of each transition on the current state.
Moreover, \acp{SMC} capture correlations between events;
for example, an interrupt that is frequently followed by a cache refill penalty will manifest as a state transition characterised by a heavier-tailed duration distribution.
This explicit, state-dependent modelling fills gaps in existing probabilistic timing analyses, offering a more interpretable framework.
Consequently, developers can readily identify which phases---such as \enquote{waiting for I/O} or \enquote{running without preemption}---contribute most to variability and worst-case latency, rather than relying solely on abstract statistical parameters.

A critical challenge in our approach entails balancing overfitting and overestimating worst-case extremes, employing overly conservative distributions.
An overfitted model represents generalisation of the system and yields inaccurate predictions, especially in data-sparse regions like the tail regions.
Our approach addresses this trade-off.
It integrates empirical measurements with a flexible, state-based stochastic framework.
This framework robustly extrapolates tail latencies.

In the following, we elaborate on the process of measurement, the incorporation of \acp{SMC}, and the construction of the stochastic model based on the recored data.

\subsection{Measurement Process}

During the measurement process, trace data is generated, which is later supplied to the generation process of the \ac{SMC} models.
Lightweight instrumentation techniques capture precise timestamps and contextual metadata from relevant system events, ensuring close alignment between observations and the actual system state.

The workflow begins with the setup of instrumentation mechanisms, including kernel tracepoints, user-space logging, and manual instrumentation.
Domain experts select events to be traced based on their understanding of system internals, the structure and implementation of real-time tasks, and potential sources of interference, such as task switches, interrupt handling, or function entry and exit.
Where tracing spans domain boundaries (\eg, kernel and user space), time consistency is achieved by using a shared monotonic clock.

During execution of a defined scenario---often repeated under varying side conditions---the system records events in chronological order.
Each event is annotated with a high-resolution timestamp, a unique identifier, and optional context (\eg, CPU or process ID) used to differentiate concurrent execution contexts.

Following data collection, the dataset is examined for anomalies, including duplicate timestamps, missing entries, and domain desynchronisation.
Filtering and consistency checks are then applied to ensure the integrity and fidelity of the captured temporal behaviour.

\subsection{Stochastic Model}

We model the runtime behaviour of a task with an \ac{SMC}.
In our formulation, the task's execution time (from initiation to completion) equals the \ac{SMC}'s time-to-absorption.
We represent the segmented execution structure by a stochastic sequence of states, where hold-times correspond to transition durations.
System events (\eg, interrupts, signals, exceptions, tracepoints, and function entry/exit points) link the observed system to the model by serving as the nodes of the \acl{SMC}.
Each realisation of the stochastic process represents one task execution instance, and its time-to-absorption directly reflects its latency.
The estimated \ac{WCET} $\hat{T}_{\max} = \max \left\{ T_{\text{abs}} \right\}$ is therefore given by the maximal time-to-absorption $T_{\text{abs}}$.
We assume that all executions reach an absorbing state in finite time, excluding non-terminating paths.
While loops may occur, the probability of infinite recurrence is assumed to be zero.

Determining $\hat{T}_{\max}$ analytically proves challenging for complex models.
We instead simulate the process by sampling iteratively, following the probability for next-step transitions $(J_n)_{n \in \mathbb{N}}$ and the sojourn times $(S_n)_{n \in \mathbb{N}}$ that match the hold-times.
For a dataset of $n$ simulation runs, we define
$\hat{T}_{\max} = \max \left\{ \sum_{i=0}^n T_i \right\}$.

Although the events' timestamps are inherently discrete---due to finite resolution of hardware timers---we adopt a continuous index set $\mathcal{T} \subseteq \mathbb{R}_+$.
This choice simplifies the analysis over large discrete domains and preserves the operational abstraction of a continuous time.

We model hold times between states as univariate random variables and represent their distributions using \acp{GMM}.
This approach balances flexibility and tractability, allowing the approximation of complex, potentially multimodal distributions while maintaining a manageable number of parameters.

Alternative parametric distributions---such as uniform, normal, or gamma---can also be employed;
however, they impose structural assumptions on the shape of the data and are often insufficient to capture irregular or heavy-tailed behaviours.
In contrast, \acp{GMM} offer a data-driven, non-restrictive alternative that supports modelling across a wide range of empirical scenarios.
Our emphasis is therefore on maximising expressive power rather than enforcing specific distributional forms.

\subsection{Model-Construction}

\begin{table}
  \caption{Mapping between model components and measurement data}
  \centering
  \begin{tblr}{
      colspec = {cl},
      rowhead=1,
      row{1}={font=\scshape,bg=lightgray},
      column{1}={mode=dmath},
      cell{1}{1}={mode=text},
    }
    \toprule
    Model & Measurement \\\midrule
    \SetRow{font=\bfseries} \SetCell[c=2]{l} \emph{Transition-Level Mapping} & \\
    \mathcal{Q} & Event Set \\
    \mathcal{S} \subset \mathcal{Q} & Start Event(s) \\
    \mathcal{A} \subset \mathcal{Q} & Termination Event(s) \\
    P & Event Sequence \\
    Q & Elapsed Time Between Events \\
    \pi & Frequency of Start Event(s) \\
    \midrule
    \SetRow{font=\bfseries} \SetCell[c=2]{l} \emph{Derived Properties} & \\
    T_{\text{abs}} & Task Duration \\
    T_{\max} & WCET \\\bottomrule
  \end{tblr}
  \label{table:mapping}
\end{table}

We derive a \acl{SMC} model from the runtime traces comprising timestamps, event identifiers, and optional context.
\autoref{table:mapping} summarises the semantic correspondence between measurement data and the model components, distinguishing transition-level elements from derived properties.

The recorded events determine the state space~$\mathcal{Q}$.
Based on our instrumentation semantics, we manually specify the set of initial states~$\mathcal{S}$ and absorbing states~$\mathcal{A}$, thereby delineate the system's execution paths present in the dataset.

The empirical initial distribution~$\hat\pi$ is estimated as the relative frequency of initial states across~$N$ observed executions
$\hat\pi_i = \nolinebreak {\#_{runs}(J_0 = q_i)} / {N}$,
where $\#_{runs}(\cdots)$ denotes the number of runs satisfying the given condition.

Transition probabilities are inferred from observed transitions.
The empirical probability of transitioning from $q_i$ to $q_j$ is
\begin{align*}
  \hat P_{ij} = \frac{\#_{trans}(q_i \to q_j)}{\#_{trans}(q_i \to \ast)},
\end{align*}
with $\#_{trans}(q_i \to q_j)$ counting transitions from $q_i$ to $q_j$,
and $\#_{trans}(q_i \to \ast)$ all transitions originating in $q_i$.
This formulation ensures the matrix $\hat P$ is row-stochastic.

Hold times are modelled using \acp{GMM}, fitted via \acl{EM} to match the timestamp differences between source and target events of each transition.
For a transition from $q_i$ to $q_j$, the hold time distribution is expressed as
$F_{ij}(t) = \nolinebreak \sum_{k=1}^{K} \omega_k \mathcal{N}(t \mid \mu_k, \sigma_k^2)$,
where $\omega_k$, $\mu_k$, and $\sigma_k^2$ denote the weight, mean, and variance of the $k$-th component, respectively, and $K$ is the number of mixture components.
To ensure plausibility, the distribution is truncated at $t=0$ and normalised, such that negative durations are excluded and $F_{ij}(t)$ is defined only for $t \geq 0$.

While standard model selection techniques---such as the \ac{AIC} or \ac{BIC}---may be employed to determine a suitable value of $K$, in our setting the number of components is typically chosen by the developer or domain expert based on the observed complexity of the empirical distribution.

\subsection{Conceptual Demonstration of Application}%
\label{sec:demo}

\begin{figure}
  \centering
  \includegraphics{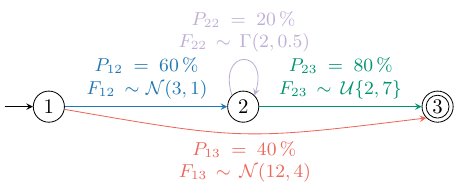}
  \caption{
    Generator model for synthetic dataset.
    Starting from the initial state $q_1$ the model transitions to $q_2$ in \SI{60}{\percent} and directly to the absorbing state $q_3$ in \SI{40}{\percent}.
    Both transition durations are normally distributed $\mathcal{N}(\mu, \sigma^2)$.
    The model contains a loop at $q_2$ which is taken in \SI{20}{\percent} with the latency distribution following a gamma distribution $\Gamma(\alpha, \lambda)$.
    The transition from $q_2$ to $q_3$ is uniformly distributed $\mathcal{U}\{a, b\}$.
  }
  \label{fig:demo:model}
\end{figure}

To illustrate our method, we apply it to a synthetic example resembling typical real-time task behaviour, where latency spans from a trigger event to a termination condition.
We generate a dataset of approximately \num{2700} event-timestamp pairs from \num{1000} independent executions of the generator model shown in \autoref{fig:demo:model}, using predefined (assumed base truth) transition probabilities and sampling transition durations from normal, gamma, or uniform distributions with a random noise value added to emulate realistic measurement conditions and to capture varying characteristics of different execution segments.
In the modelling phase, all hold-times---regardless of their origin---are re-approximated using \acp{GMM}.
Distributions with support on negative time ($\Pr(t < 0) > 0$) are truncated at $t = \epsilon > 0$ during sampling to enforce non-negativity.

We partition the generated event log into individual runs using $q_1$ as the initial and $q_3$ as the absorbing state with $\pi^{(0)}_1=1$.
From this, \num{24} independent \ac{SMC} models over $\mathcal{Q} = \{ q_1, q_2, q_3 \}$ are reconstructed.
Transition probabilities are inferred from sequences in the data, and holding times modeled via \acp{GMM}.
To estimate the overall latency distribution, each model is simulated \num{2000} times.
The number of \acs{SMC} models and the number of runs reflects a trade-off between computational cost and statistical robustness, capturing variability due to stochastic fitting convergence and enabling parallelised evaluation.

\begin{figure}
  \centering
  \includegraphics{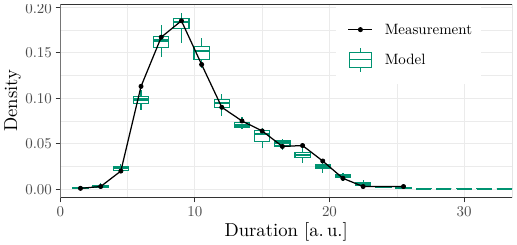}
  \caption{
    Comparison of latency distributions for the synthetic dataset for empirical measurements and model-based analysis.
    The black frequency polygon represents the latency distribution (from initiation to termination) derived from the synthetic dataset, while the green box plots show the aggregated density of latencies for each duration bin, as modeled by \num{24} independent \aclp{SMC}.
  }
  \label{fig:demo:distribution}
\end{figure}

\autoref{fig:demo:distribution} demonstrates the close match between empirical and simulated latency distributions, confirming that our method reliably reconstructs timing behaviour under noise.
We revisit this example in Sections~\ref{sec:evaluation} and~\ref{sec:discussion} to analyse tail quantiles and maximum values for \ac{pWCET} estimation.

\section{Implementation Components}%
\label{sec:implementation}

Our method requires three basic components:
(A)~tracepoints as data sources of events captured during runtime,
(B)~a tracing tools to collect and store events, and
(C)~a data analysis framework that processes the recorded data, generates an \ac{SMC} model and evaluate its predictions.

The following describes the design and implementation of our custom instrumentation and event tracing mechanism, followed by a brief description of our analysis framework.

\subsection{Timed Tracepoints}
\label{sec:ttp}

To collect timing information with minimal runtime interference, we introduce a custom lightweight tracing mechanism termed \emph{Timed Tracepoints}~(TTPs).
Conventional tracing frameworks such as \texttt{ftrace} or \texttt{eBPF}, although expressive and widely supported, may introduce non-negligible overheads that compromise the precision of temporal measurements.
To overcome these limitations, \acp{TTP} are designed to record only essential data:
a unique event identifier,
a high-resolution timestamp,
and minimal contextual information (\eg, \ac{CPU} or process \ac{ID}).
This reduction in scope ensures significantly lower overhead while retaining sufficient detail for probabilistic timing analysis.

Each \ac{TTP} consists of a static probe, that can be inserted across the kernel, and a global handler, statically compiled to avoid dynamic instrumentation costs.
Trace events are logged into a statically allocated memory buffer using a global interface function (\texttt{ttp\_emit}), thereby eliminating the need for dynamic memory allocation and preventing timing perturbations during logging.
The interface exposes only a minimal control surface---namely enable, disable, and reset operations---to further reduce runtime impact.
After system execution, the contents of the buffer are retrieved from user space and processed in an offline analysis step.

\subsection{Data Analysis Framework}

The model construction and statistical evaluation is implemented in R~\cite{r_core_team_r_2024}, and leverages its ecosystem of parallelised data manipulation, statistical analysis, and visualisation libraries~\cite{wickham_welcome_2019,wickham_ggplot2_2016,barrett_datatable_2024}.
The core workflow integrates data preprocessing, transformation, and statistical modelling to construct the \ac{SMC} model and analyse system runtime behaviour.
The complete codebase is published as open-source software.%
\footnote{We published our code and data at \url{https://github.com/lfd/smc-ta}.}

We generate the \acl{SMC} model and its estimation by processing measurement data in a simple \ac{CSV} format.
First, we transform raw event data into a time-ordered log of transitions that correspond to individual task  runs.
We scale timestamps to a consistent unit for accuracy and compatibility.
We then apply filters to discard incomplete runs and exclude irrelevant events occurring outside the span of the designated start and termination events.
Additional checks verify assumptions about the data, such as unique timestamps within each execution run.

Next, we construct the \ac{SMC} from the refined transition data.
We compute transition probabilities and derive transition duration distributions modelled with \acp{GMM}, which we fit using the \acl{EM} algorithm.

Finally, we perform simulation runs to sample the distribution of \acl{ET} from the constructed \ac{SMC}.
We generate random samples of the \ac{SMC}'s trajectories to estimate runtime characteristics of the system, including the \acl{WCET}.
These simulations account for variability in both transition probabilities and sojourn times, yielding probabilistic bounds for runtime analysis.

\section{Evaluation}%
\label{sec:evaluation}

In the following, we demonstrate our method with cyclictest\footnote{Cyclictest is part of \texttt{rt-tests}, a suite of tools for evaluating real-time behaviour in Linux: \url{https://git.kernel.org/pub/scm/utils/rt-tests/rt-tests.git}.}, a widely used benchmarking utility for measuring scheduling latencies in real-time Linux systems.
It reports latency as the deviation between a timer's programmed wake-up time and the actual time at which the corresponding task is resumed.
While this notion of latency differs from classical execution time---typically defined as the duration from task release to completion---the distinction is immaterial for our approach, which focuses on the temporal distance between a defined start and termination event, irrespective of semantic interpretation.

We adopt cyclictest for three reasons:
(i) it exercises the full stack from kernel interrupt handling to user-space scheduling;
(ii) it is well understood by kernel developers, including the meaning and structure of its reported jitter;
and (iii) it allows us to quantify the instrumentation overhead by comparing latency distributions of annotated and unannotated systems.

In our model, the expected wake-up time is treated as the start event and the actual wake-up time as the absorbing state.
The following sections describe the experimental setup, measurement procedure, derived results, and an analysis of robustness under partial data conditions.

\subsection{Experimental Setup}

We conduct measurements on a \emph{StarFive VisionFive~3}, a \acl{SBC} featuring a \SI[unit-mode=text]{64}{\bit} RISC\nobreakdashes-V processor.
It runs a real-time enabled Linux kernel (\texttt{6.11.0-rc5}), configured with the \texttt{PREEMPT\_RT} patch stack.
Any unused devices and functionalities are disabled. To further RT-tune the system, system noise is reduced by the isolation of \acp{CPU} via boot parameter to pin real-time tasks to specific cores.

To minimise latency perturbations while preserving high precision, the kernel was patched with our lightweight \acp{TTP}.%
\footnote{Our \acs{TTP} implementation and instrumentation of the Linux kernel is available at \url{https://github.com/lfd/linux/tree/smc-ta-ttp}.}
Tracepoints are inserted at critical locations targeting the key sources of latency in our evaluation:
the \texttt{riscv\_timer\_interrupt} event handler (capturing entry~\circledinline{2} and exit~\circledinline{3}) and within the scheduler to record context switches~\circledinline{1} (\texttt{TRACE\_SCHED\_SWITCH}) as well as task wake-ups~\circledinline{4}\,\circledinline{5} (\texttt{TRACE\_SCHED\_WAKEUP} and \texttt{TRACE\_SCHED\_WAKING}).
To capture expected and actual wake-up times, two additional \acp{TTP} are integrated into the cyclictest application using the same time base as the ones in the kernel.
Data from both the kernel and the application were recorded independently and later merged for offline analysis on a secondary machine.

Cyclictest is executed on the real-time Linux system with a real-time priority of \num{90} and \ac{FIFO} scheduling to ensure deterministic behaviour.
It was configured to capture \num{1000} measurements per second over a period of \SI{5}{\min} yielding a dataset of \num{600}k individual runs.
The worst-case latency observed during our measurement is \SI{51.58}{\micro\second}.

\begin{figure}
  \centering
  \includegraphics{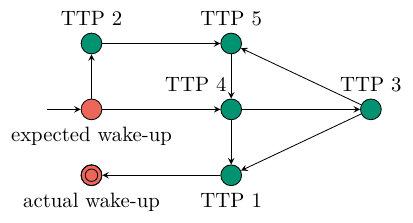}
  \caption[State Model of \ac{SMC} for cyclictest]{
    State model of a \acl{SMC} for cyclictest.
    The model comprises states for each tracing events.
    Two states (red) represent the expected (initial state) actual wake-up (absorbing state) within the cyclictest application.
    The remaining states (green) correspond to the manually inserted \acl{TTP} in the kernel's interrupts handler and scheduler.
  }
  \label{fig:cyclictest:model}
\end{figure}

\subsection{Results}

To capture the variability inherent in both system execution and probabilistic modelling, we generate \num{24} independent \ac{SMC} instances from the recorded event data.
This ensemble-based approach ensures that model construction is robust to stochastic effects in both data and parameter estimation.
Each \ac{SMC} encodes transition durations using a \ac{GMM} with four components.
This number was empirically chosen as a trade-off between model flexibility and overfitting, providing sufficient expressiveness to approximate multi-modal timing behaviours without introducing excessive estimation variance or computational cost.
\autoref{fig:cyclictest:model} illustrates a simplified representation of the resulting state structure and transitions, reflecting the temporal ordering of key events, including both kernel-level \acp{TTP} and cyclictest measurements.

For each \ac{SMC}, we conduct \num{10} independent Monte Carlo simulations to propagate timing uncertainty and sample the system's latency distribution.
Each simulation dataset comprises \num{10000} absorption runs---complete state sequences from initial to terminal states---capturing the variability in both path selection and transition timing.
This sample size aims for statistical stability in tail estimation while remaining computationally tractable.
We analyse each resulting distribution regarding various quantiles and average its maximal value over the \num{10} simulation runs.

\begin{figure}
  \centering
  \includegraphics{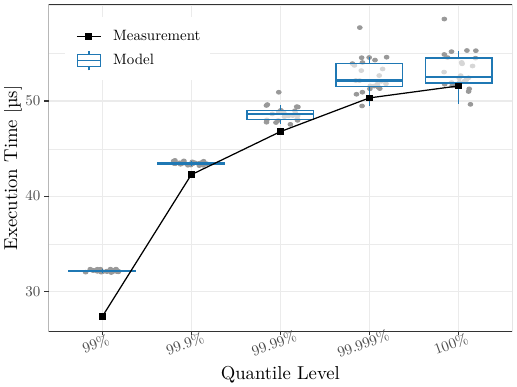}
  \caption[Quantile Comparison for Cyclictest Case Study]{%
    Comparison of quantiles of execution times from the measurement (black squares) and model sampling simulation (blue), aggregated over \num{24} independently built semi-Markov chains.
    The connecting line is included only to guide the eye and does not represent interpolation or progression.
  }
  \label{fig:cyclictest:results}
\end{figure}

\autoref{fig:cyclictest:results} compares the quantiles obtained from  empirical measurements (red) with those derived from the model simulation data (blue).
The mean of the models' \ac{WCET} estimation is \SI{53.14}{\micro\second}, which overestimates the empiric \ac{WCET} by about \SI{3}{\percent}.
The quantile values averaged over the \num{24} generated \acp{SMC} exhibit overestimation of \SI{4.7}{\percent} for the \SI{99.999}{\percent} quantile, \SI{4.0}{\percent} for the \SI{99.99}{\percent}, and \SI{2.9}{\percent} for \SI{99.9}{\percent} quantile.

The results indicate a strong correspondence between the model-derived quantiles and the values observed in the original dataset, particularly up to the \SI{99.9}{\percent} quantile.
Although the variability increases for the \SI{99.99}{\percent} quantile and the estimated worst-case latency---as reflected by the box plots---the mean and median values across simulations remain very close to those measured.
These findings suggest that our approach reliably captures the latency distribution over a wide range of quantiles, while the increased uncertainty in the tail reflects the inherent variability in rare, extreme events.

\subsection{Model Robustness with Partial Datasets}

\begin{figure*}
  \centering
  \includegraphics{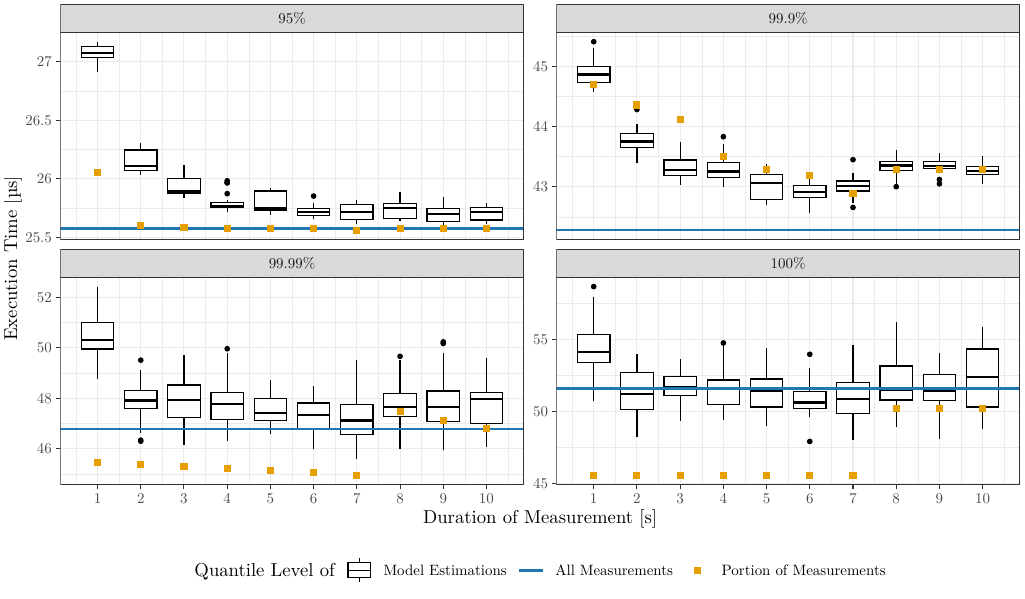}
  \caption{%
    Convergence of quantile estimates with reduced measurement duration.
    Quantile estimates are derived from simulation data generated using subsets of the full dataset.
    For each measurement duration (first $x$\,seconds, \ie, $x \times 10000$ runs), \num{24} SMC models are generated.
    Each of the models is sampled ten times with \num{10000} runs.
    The models' prediction is the mean over quantiles calculated for each of the \num{10} sampling runs.
    Black box plots show the quantile predictions from \ac{SMC} models, golden squares represent measured quantiles, and the blue line represents the quantile computed from the full dataset.
  }
  \label{fig:cyclictest:tradeoff}
\end{figure*}

To assess the impact of limited measurement data on model accuracy, we generated partial datasets by selecting only the first $x$ seconds (\ie, $x \times 1000$~runs) from the full dataset.
For each subset, we follow the same analysis procedure:
independently build \num{24} \acp{SMC}, generate \num{10} sampling datasets with \num{10000} simulation runs each.

\autoref{fig:cyclictest:tradeoff} compares the model predictions (black box plots) with the corresponding measured quantile values from the partial datasets (golden dots) and with the quantiles computed from the full dataset (blue line).
The figure clearly shows that, even with a substantially reduced dataset, the \ac{SMC}-based predictions converge toward the reference values from the full dataset.
Notably, even when we reduce the measurement duration to only a few seconds, the model's predictions for extreme quantiles (\eg, the \SI{99.99}{\percent} quantile) and the \ac{WCET} remain stable.
For instance, when we use a \num{2}\,second subset---representing only \SI{0.3}{\percent} of the full \SI{5}{\minute} dataset---the empirical \ac{WCET} in this subset is \SI{45.56}{\micro\second}, which is lower than the \SI{51.16}{\micro\second} present in the complete dataset.
Despite the absence of the latter in the subset, the averaged prediction from the models approximates the full-dataset \ac{WCET}, yielding \SI{51.16}{\micro\second}.

The impact of data reduction is more apparent at lower quantiles.
Convergence of the \SI{95}{\percent}~quantile emerges only after approximately \num{6}~seconds of measurement.
In contrast, the predictions for the \SI{99.99}{\percent}~quantile and the \ac{WCET} stabilise after \num{2}~seconds and undergo only minor variance with longer measurement durations.
On average, the models slightly overestimate the quantile values relative to both the full dataset and the subsets, except for the \SI{99}{\percent} quantile.
We observe a sudden increase in the \SI{99.99}{\percent} quantile between \numlist{7;8}\,seconds, which likely reflects a higher incidence of elevated latency values in the smaller datasets.
As more data becomes available, the quantile estimate gradually converges towards the value derived from the full dataset.

\section{Discussion}%
\label{sec:discussion}

The evaluation demonstrates that our method effectively models the timing behaviour of cyclictest latency via \acp{SMC}.
The event-centric approach integrates user space and kernel level data, ensuring that both frequent and rare extreme behaviours receive accurate representation.
The complexity of \ac{SMC} models scales with the number of distinct events and observed transitions rather than with the number of execution paths.
This scaling avoids the exponential complexity that plagues many static analysis techniques~\cite{davis_survey_2019}.
The method provides significant advantage for complex, real-world systems, as it requires only a log of events with precise timestamps.

The robustness analysis suggests that our approach effectively generalises from limited data.
Even when we use a few seconds of measurement rather than minutes, the \ac{SMC} models extrapolate and predict extreme execution times despite the absence of those extremes in the measurement window.
This reduction in required measurement time and resources proves especially beneficial when such resources are constrained.
Rapid feedback on system performance accelerates the development cycle.
Furthermore, the minor conservative bias---where the model tends to overestimate lower quantiles slightly---can serves as an advantage and ensures that predicted worst-case scenarios remain on the safe side.

We observe a rise in the \SI{99.99}{\percent} quantile between \num{7} and \num{8} seconds, which indicates sensitivity in the estimation of extreme events.
This observation highlights an area for further refinement.
Future work may enhance the model's robustness by incorporating adaptive weighting mechanisms or context-sensitive constraints to more accurately capture such transitions.
Overall, these findings reinforce the value of our hybrid method, which provides dependable runtime predictions and offers runtime insights for further system optimisation.

We applied our method to an alternative test operating system using the same approach.
We instrumented the kernel and implemented an application analogous to \texttt{cyclictest}.
Results agree with those obtained for the original system.

Our framework provides an intuitive and interpretable representation of system latency by modelling execution time as time-to-absorption of the \ac{SMC}.
It aids developers to identify key characteristics---such as waiting for I/O or interrupt handling---that contribute most significantly to performance variability.

The reliance on \acp{GMM} to model transition durations risks overfitting, particularly in scenarios with low variability.
We also encountered logically invalid state sequences in our experiments.
Incorporating context-sensitive constraints could prevent infeasible state transitions and enhance model validity.
These issues are not unique to our approach, however;
other measurement-based timing analysis methods also suffer from overfitting and incomplete representation of relevant factors.

The effectiveness of our method depends on the quality and granularity of both the user-provided model (through instrumentation) and the measurement data.
The model's predictive accuracy relies on matching its complexity with that of the underlying system.
Overly simplistic models may omit critical execution paths, while overly complex models risk overfitting the available data.
A limitation to our current implementation is the assumption of time-homogeneity.
We also simplify by assuming that transition durations depend solely on the source and target states, thereby neglecting possible long-term dependencies and context-sensitive effects, such as whether preemption is disabled.

Although our demonstration confirms the model's generalisation capability, challenges remain.
Missing transitions yield an incomplete state model that may reduce accuracy~\cite{lesage_framework_2015}.
The model's simplicity, however, permits mitigation.
We can address missing transitions and adjust assumptions regarding the distribution of transition durations between events manually.
In such cases, we readily adapt the model by modifying its probabilistic parameters, such as the transition probability matrix $P$ or the holding-time kernel $Q$.

\section{Conclusion}%
\label{sec:conclusion}

We presented a semiformal hybrid method for runtime analysis that integrates measurement-based observations with model-based techniques.
We employ \aclp{SMC} to model execution times (latency between critical system events).
Our approach balances realism and statistical precision, while preserving simplicity.
It operates at a high level of abstraction, and primarily requires a choice of statistical distribution for transition durations.
Our method requires only minor modifications to the target system, such as enabling and configuring lightweight tracepoints.
As demonstrated in our evaluation, the approach integrates runtime interference within and across different domains, while imposing minimal overhead without compromising accuracy.

Our \ac{SMC} models capture the structural characteristics of task execution through inter-event latencies.
The experimental evaluation with cyclictest shows that the approach overestimates the empirical \ac{WCET} by only about \SI{3}{\percent} on average.
This level of precision persists even when the measurement duration is reduced from multiple minutes to only a few seconds.
Although longer measurements are typically needed to capture worst-case events directly, our framework predicts \ac{WCET} from a limited dataset.
This offers developers rapid insights into system timing behaviour, enabling quick feedback on the impact of code modifications and accelerating development.

In summary, our approach bridges rigorous modelling and pragmatic measurement-driven validation.
It combines empirical data with stochastic models, and is applicable to rapid prototyping and in-depth performance evaluation.

We plan to extend the framework to support online analysis and integrate it with established real-time monitoring tools.
This will enable real-time verification, adaptive instrumentation, and broader applicability in dynamic and industrial settings.

On the modelling side, we aim to incorporate context-sensitive factors (\eg, temperature, voltage, preemption state) to improve predictive accuracy and responsiveness to system conditions.
We will also explore richer sojourn time distributions to better capture tail behaviour and rare-event dynamics.

Finally, we plan to apply the method to larger and more complex systems, including multicore and distributed real-time platforms, to evaluate scalability and demonstrate its utility in realistic deployment scenarios.

\printbibliography

\end{document}